\documentclass[aps,reprint,superscriptaddress]{revtex4-2}
\usepackage{siunitx}
\usepackage{graphicx}
\usepackage{placeins}
\usepackage{textcomp}
\usepackage{amssymb}
\usepackage{makecell}
\usepackage{physics}

\begin{document}

\title{Cavity-enhanced single-shot readout of a quantum dot spin within 3\,nanoseconds}

\author{Nadia O. Antoniadis}
\thanks{These two authors contributed equally}
\affiliation{Department of Physics, University of Basel, Klingelbergstrasse 82, CH-4056 Basel, Switzerland}
\author{Mark R. Hogg}
\thanks{These two authors contributed equally}
\affiliation{Department of Physics, University of Basel, Klingelbergstrasse 82, CH-4056 Basel, Switzerland}
\author{Willy F. Stehl}
\affiliation{Department of Physics, University of Basel, Klingelbergstrasse 82, CH-4056 Basel, Switzerland}
\author{Alisa Javadi}
\affiliation{Department of Physics, University of Basel, Klingelbergstrasse 82, CH-4056 Basel, Switzerland}
\author{Natasha Tomm}
\affiliation{Department of Physics, University of Basel, Klingelbergstrasse 82, CH-4056 Basel, Switzerland}
\author{R\"{u}diger Schott}
\affiliation{Lehrstuhl f\"{u}r Angewandte Festk\"{o}rperphysik, Ruhr-Universit\"{a}t Bochum, D-44780 Bochum, Germany}
\author{Sascha R. Valentin}
\affiliation{Lehrstuhl f\"{u}r Angewandte Festk\"{o}rperphysik, Ruhr-Universit\"{a}t Bochum, D-44780 Bochum, Germany}
\author{Andreas D. Wieck}
\affiliation{Lehrstuhl f\"{u}r Angewandte Festk\"{o}rperphysik, Ruhr-Universit\"{a}t Bochum, D-44780 Bochum, Germany}
\author{Arne Ludwig}
\affiliation{Lehrstuhl f\"{u}r Angewandte Festk\"{o}rperphysik, Ruhr-Universit\"{a}t Bochum, D-44780 Bochum, Germany}
\author{Richard J. Warburton}
\email[To whom correspondence should be addressed:]{ mark.hogg@unibas.ch, richard.warburton@unibas.ch}
\affiliation{Department of Physics, University of Basel, Klingelbergstrasse 82, CH-4056 Basel, Switzerland}

\date{\today}

\begin{abstract}
	
Rapid, high-fidelity single-shot readout of quantum states is a ubiquitous requirement in quantum information technologies, playing a crucial role in quantum computation, quantum error correction, and fundamental tests of non-locality. Readout of the spin state of an optically active emitter can be achieved by driving a spin-preserving optical transition and detecting the emitted photons. The speed and fidelity of this approach is typically limited by a combination of low photon collection rates and measurement back-action. 
Here, we demonstrate single-shot optical readout of a semiconductor quantum dot spin state, achieving a readout time of only a few nanoseconds.
In our approach, gated semiconductor quantum dots are embedded in an open microcavity. 
The Purcell enhancement generated by the microcavity increases the photon creation rate from one spin state but not from the other, as well as efficiently channelling the photons into a well-defined detection mode. 
We achieve single-shot readout of an electron spin state in 3\,nanoseconds with a fidelity of (95.2$\pm$0.7)\%, and observe quantum jumps using repeated single-shot measurements. Owing to the speed of our readout, errors resulting from measurement-induced back-action have minimal impact. Our work reduces the spin readout-time to values well below both the achievable spin relaxation and dephasing times in semiconductor quantum dots, opening up new possibilities for their use in quantum technologies.
	
\end{abstract}

\maketitle

\section{Introduction}
The ability to perform a projective measurement of a quantum state in a single measurement (single-shot readout) is an enabling technique in quantum technologies \cite{Elzerman2004,Neumann2010}. Single-shot readout is necessary in quantum computation in order to extract information at the end of the protocol, as well as in error detection and correction as the quantum processor runs \cite{Fowler2012}. Additionally, single-shot readout is necessary to close the fair-sampling loophole in tests of quantum non-locality, and was a key ingredient in recent demonstrations of loophole-free Bell inequality violations \cite{Hensen2015}. 
The ideal single-shot readout protocol achieves high-fidelity qubit readout in the shortest time possible; readout within the qubit dephasing time is essential for quantum error correction, and enables measurement-based quantum feedback \cite{Vijay2012, Campagne2013} and quantum trajectory tracking \cite{Murch2013}.

The spin states of semiconductor quantum dots (QDs) show exceptional promise in quantum technology \cite{Watson2018, Hendrickx2021, Philips2022}. Optically-active QDs, established bright and fast sources of coherent single photons \cite{Somaschi2016,Wang2019PRL,Uppu2020,Tomm2021}, can be occupied with a single electron and the electron spin can be initialised \cite{Atature2006,Xu2007} and rotated on the Bloch sphere \cite{Press2008, Bodey2019} on nanosecond timescales using all-optical techniques. Theoretical proposals \cite{Lindner2009, Tiurev2020} and recent experiments \cite{Schwartz2016, Cogan2021arXiv, Coste2022arXiv} have established the spin-photon interface provided by the InGaAs platform as a leading contender for creating photonic cluster states, an important resource for quantum repeaters \cite{Borregaard2020} and measurement-based quantum computation \cite{Walther2005}. The dephasing time of the electron spin in optically-active QDs is limited by magnetic noise arising from the nuclear spins. However, there are powerful mitigating strategies. A double-QD can be used to create a clock-transition \cite{Weiss2012}; a switch to a hole spin suppresses the effect of the magnetic noise particularly in an in-plane magnetic field \cite{Brunner2009,Prechtel2016}; and the noise can be almost eliminated by laser-cooling the nuclei \cite{Gangloff2019,Jackson2022}. In the context of cluster states, spin readout is necessary in order to disentangle the spin from the photons, thereby releasing an entirely photonic entangled state. To date, single-shot spin readout on a timescale comparable to the rapid spin initialisation and manipulation times has remained elusive.

Spin readout with an optical technique typically proceeds by applying a magnetic field to a QD containing a single electron, resonantly driving one of the Zeeman-split trion transitions, then collecting the spin-dependent resonance fluorescence \cite{Lu2010}. However, during readout, the applied laser can induce an unwanted spin flip \cite{Lochner2020,Mannel2021}, a process known as `back-action'. The key challenge for spin readout is to collect enough photons to determine reliably the spin state before the back-action flips the spin. Of the small number of previous experiments to achieve single-shot readout of InGaAs QD spin states \cite{Vamivakas2010, Delteil2014, Gillard2022}, the most rapid to date achieved a fidelity of 82\% in a readout time of 800\,ns \cite{Delteil2014}. This 800\,ns readout time was similar to the back-action timescale, and is significantly longer than the dephasing time for an electron spin bound to an InGaAs QD ($T_{2}^{*}=125$\,ns following nuclear bath cooling \cite{Gangloff2019,Jackson2022}).

In this work, we report nanosecond-timescale, all-optical, single-shot spin readout. We use an open microcavity to boost the photon collection efficiency in order to reduce the spin readout time. We achieve single-shot readout in only 3\,nanoseconds with a fidelity of (95.5$\pm$0.7)\%, an improvement in readout speed of more than two orders of magnitude with respect to previous experiments. To the best of our knowledge, this is the fastest single-shot readout of a quantum state ever achieved across any material platform. Our approach brings the readout time well below the dephasing time for an electron spin in this system. Our open microcavity approach can be used to enhance optical spin readout in other systems, such as nitrogen vacancy centres in diamond \cite{Riedel2017}.

\section{Results}
\subsection{High efficiency photon collection}

\begin{figure}[t]
	\centering
	\includegraphics[width=1\columnwidth]{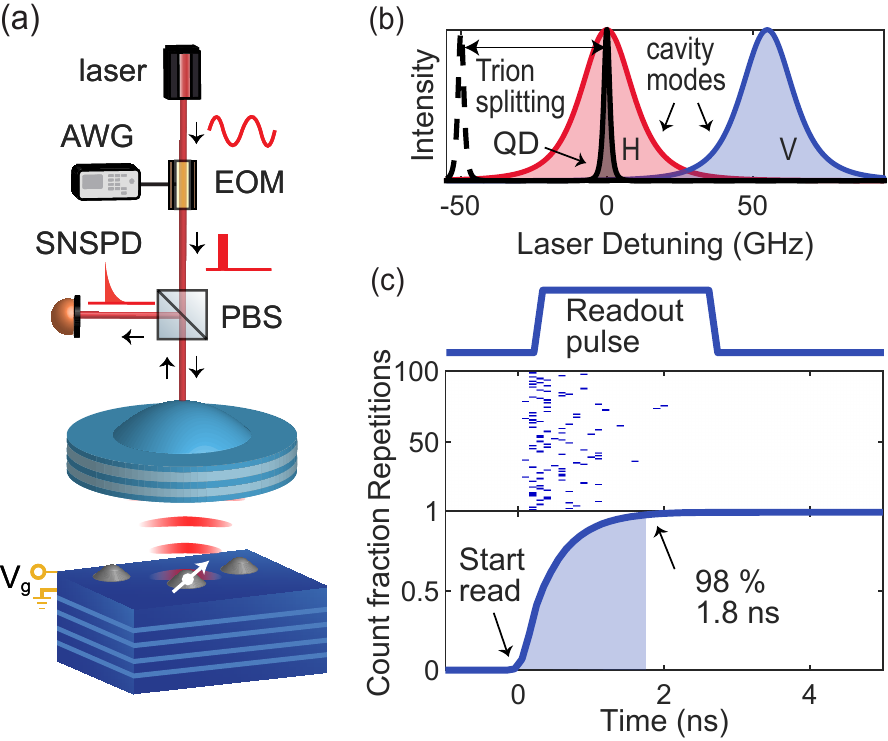}
	\caption{\textbf{Experimental setup and system efficiency.} \textbf{(a)} Resonant laser pulses with variable intensity and duration are sent to the QD using an electro-optic modulator (EOM) driven by a fast arbitrary waveform generator (AWG). 
	The photons emitted by the QD are collected in the output arm of the cross-polarised microscope and measured on a SNSPD (superconducting nanowire single photon detector). \textbf{(b)} Frequency configuration of the QD and mode-split cavity with respect to the laser. 
	With a 2.0\,T magnetic field, only one trion transition is resonant with the H-polarised cavity mode, resulting in spin-selective Purcell enhancement. 
	\textbf{(c)} Readout characterisation at zero magnetic field: here, the readout pulses are set to a duration of 2\,ns (top panel) with a repetition time of 100\,ns. Photons emitted by the QD are detected and the arrival times registered for 100,000 repetitions of the pulse sequence; 100 example traces are depicted in the middle panel where the blue dots represent a photon detection event. 
	In 98\% of the repetitions a photon is detected within 1.8\,ns.}
	\label{fig:Fig1}
\end{figure}

A schematic of the setup used in our experiments is shown in Fig.\,\ref{fig:Fig1}\,(a). Our sample is a gated, charge-tunable InGaAs/GaAs device, with a highly reflective Bragg mirror integrated into the semiconductor heterostructure \cite{Najer2019, Tomm2021}. The gate structure allows the charge occupancy of the QD to be set, as well as fine tuning of the emission frequency via the quantum-confined Stark effect. We operate with a single electron occupying the QD, which is our spin readout target.
A miniaturised Fabry-P\'{e}rot cavity is created between the semiconductor bottom mirror and a free-standing concave top mirror. The QD sample is attached to an XYZ nano-positioning stage. This flexibility of the open microcavity design allows the cavity to be re-positioned to address a chosen QD. Once a QD is positioned at the anti-node of the cavity field (XY positioning), its frequency can be matched to one of the QD transitions (Z positioning). 

\begin{figure*}[hbt]
	\centering
	\includegraphics[width=1\textwidth]{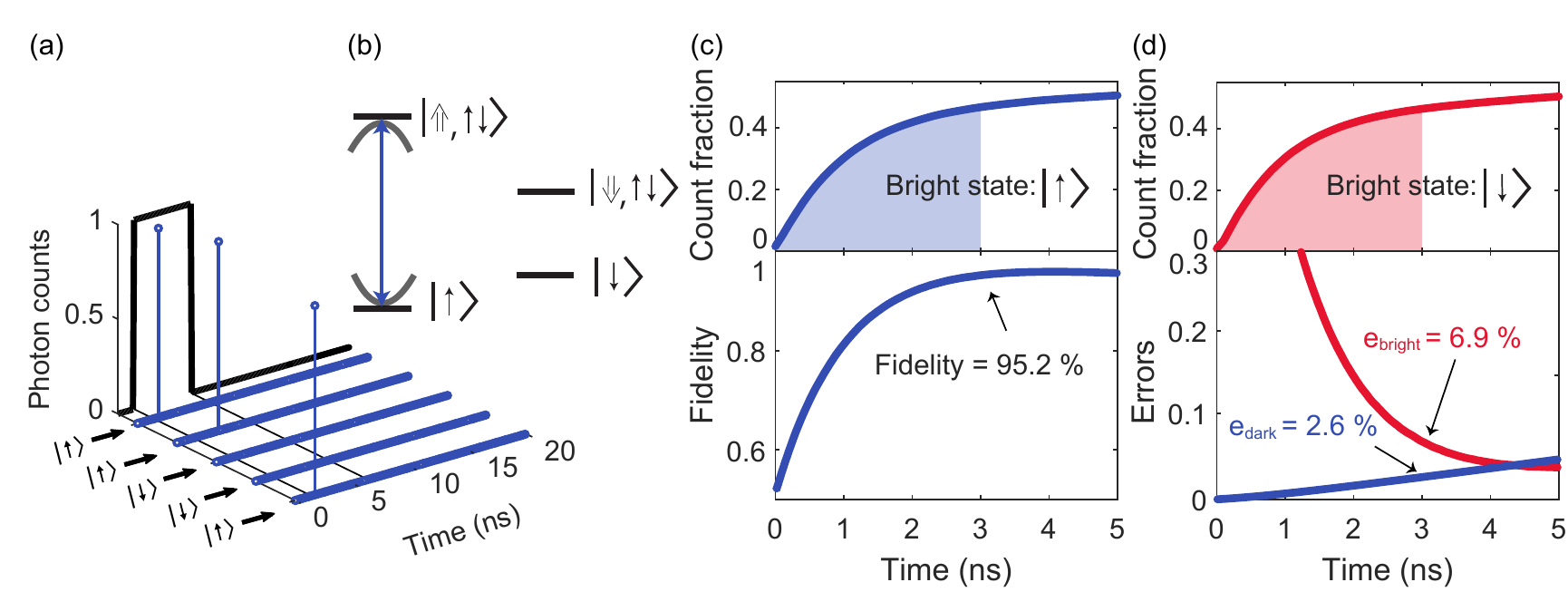}
	\caption{\textbf{Single-shot readout of the QD spin at 2.0\,T.} \textbf{(a)} Example single-shot readout traces. If a photon is detected during the readout pulse, the state of the QD is assigned to the bright state (here, spin up $\ket{\uparrow}$). Repetitions with no detected photon are assigned to the dark state (here, spin down $\ket{\downarrow}$). \textbf{(b)} Schematic of the QD energy levels in a magnetic field, indicating the readout transition (here, bright state $\ket{\uparrow}$, blue arrow) and the cavity frequency. \textbf{(c)/(d)} Experimental count fraction (top) and corresponding readout fidelity/errors (bottom) as a function of readout time for the bright state being up/down. Here, readout pulses with a duration of 5\,ns are used. The pulse sequence is repeated 100,000 times. We achieve a readout fidelity of 95.2\% for a readout time of 3\,ns.}
	\label{fig:Fig2}
\end{figure*}

Figure\,\ref{fig:Fig1}\,(c) demonstrates the high photon collection efficiency of our microcavity system and its potential for rapid spin readout. A photon emitted by the QD exits the output facet of the collection single-mode fibre with 57\% probability \cite{Tomm2021}. The overall system efficiency, $\eta$, the probability that an exciton in the QD results in a click on the detector, is 37\%. Initially, we set the magnetic field to zero, such that the optical transitions for both electron spin states are degenerate. In this scenario, a resonant laser pulse excites the QD optical transition regardless of the electron spin state. The readout pulse drives the optical transition, and the QD emits photons at a rate set by the (Purcell-enhanced) optical decay rate. The time required for a photon emitted from the QD to be registered by the detector depends on the overall system efficiency; for high efficiencies a photon is rapidly detected. 
We apply a train of 2\,ns readout pulses (temporal shape close-to-square, separated by 100\,ns, with an optical power equal to six times the QD saturation power) to the QD, and monitor the collected photons on the single photon detector (an SNSPD). The SNSPD has a dead time of $\sim12$\,ns, meaning that after one photon has been detected another detection event stemming from the same pulse is extremely unlikely. Thus, although the QD emits at a constant rate during the 2\,ns readout pulse, a maximum of one photon detection event occurs. We repeat the pulse sequence 100,000 times, and analyse the fraction of pulses in which a photon was detected as a function of the readout duration. We note that although the readout pulse length is fixed to be 2\,ns in the experiment, we can determine the fraction of pulses containing a detection event as a function of a software-defined read duration that is less than 2\,ns by analysing the photon time-of-arrival data. 
We find that for 98\% of the traces, a photon is detected within 1.8\,ns (see bottom panel Fig\,\ref{fig:Fig1}(c)). When the same pulse sequence is repeated with the QD detuned out of resonance with the readout laser, we detect a photon (due to laser leakage within the cross-polarised setup) for $<0.1\%$ of the pulses, demonstrating that the photons we detect are almost exclusively created by the QD.

\subsection{Single-shot spin readout}
To perform single-shot spin readout, we apply a magnetic field of 2.0\,T along the growth direction of the sample (Faraday configuration), which creates a four-level system in which the two strongly allowed trion transitions are split by 55\,GHz (the sum of the electron and hole Zeeman splittings, 6.8\,GHz/T and 20.7\,GHz/T respectively). Spin readout is achieved by tuning the cavity into resonance with one of the strongly allowed transitions, as shown in Fig.\,\ref{fig:Fig2}(b). The readout pulse sequence is then similar to that shown in Fig.\,\ref{fig:Fig1}(c), but photon emission is now only enhanced for the trion transition resonant with the cavity.
Figure\,\ref{fig:Fig2}\,(a) shows example single-shot readout traces: here, we apply a train of readout pulses (5\,ns duration with a repetition time of 100\,ns) resonant with the cavity-enhanced $\ket{\uparrow} \leftrightarrow \ket{\Uparrow,\uparrow\downarrow}$ trion transition. We note that for this experiment, the frequency alignment of cavity modes and trion transitions is identical to that shown in Fig.\,\ref{fig:Fig1}\,(b). 
If the electron is projected into the $\ket{\uparrow}$ spin state, Purcell-enhanced fluorescence from the $\ket{\uparrow} \leftrightarrow \ket{\Uparrow,\uparrow\downarrow}$ trion will be rapidly registered by the detector. The spin is thus projected into the ``bright" state, and detecting a single photon emitted by the QD during the readout pulse constitutes a measurement of the spin state. Conversely, if the electron is projected into the $\ket{\downarrow}$ (``dark") spin state no fluorescence is detected, as the $\ket{\downarrow} \leftrightarrow \ket{\Downarrow,\uparrow\downarrow}$ trion is out of resonance with the readout laser. In this case, the absence of a detector event during the readout pulse indicates that the spin was projected into the dark state. We stress again that the readout time is less than the dead time of the detector: a maximum of one photon can be measured during the readout process. Furthermore, the overall system efficiency is high enough that the \emph{absence} of a detected photon contains significant information: it denotes that the spin was projected into the dark state. The detection is thus binary: detection of one photon corresponds to the $\ket{\uparrow}$ state, and zero photons to the $\ket{\downarrow}$ state. Equivalently, our photon number threshold for discriminating the spin states is one single photon. 

We repeated the spin-readout measurements with the cavity and readout laser tuned such that either $\ket{\uparrow}$ or $\ket{\downarrow}$ is the bright transition. In Fig.\,\ref{fig:Fig2}\,(c) we show the results of 100,000 repetitions of the spin readout pulse sequence with $\ket{\uparrow}$ set as the bright state (the configuration shown in Fig.\,\ref{fig:Fig2}\,(b)). We plot the fraction of readout traces containing one photon, i.e.\ the fraction of traces we assign the electron spin state to be $\ket{\uparrow}$. 
We observe a rapid increase in the count fraction (on a timescale of a few nanoseconds) as a function of the readout time.
Compared to the 0\,T results in Fig.\,\ref{fig:Fig1}\,(c), the maxima of the count fractions now saturate close to 50\%: each spin state is almost equally likely. The reason is that the spin is not initialised in these experiments. Instead, before readout, the spin is in a mixed state as co-tunnelling between the QD and the Fermi sea of the back contact regularly randomises the spin state (on a timescale of $\sim 300$\,ns) during the 100,000 readout pulse repetitions, such that  both $\ket{\uparrow}$ and $\ket{\downarrow}$ spin states have approximately equal probabilities.
Figure\,\ref{fig:Fig2}\,(d) shows data for 100,000 repetitions of the readout pulse sequence, now with the readout laser resonant with the low frequency trion transition, $\ket{\downarrow} \leftrightarrow \ket{\Downarrow~\uparrow\downarrow}$ (thus making the $\ket{\downarrow}$ state the bright state and $\ket{\uparrow}$ the dark state).

Compared to the 0\,T readout in Fig.\,\ref{fig:Fig1}(c), the readout speed is slightly slower (high-fidelity readout is achieved in 3\,ns rather than 1.8\,ns). The reason for this slower readout is that at 2.0\,T we operate with the laser on resonance with the QD but detuned by 7.5\,GHz from the actual cavity resonance, where we observe optimal laser suppression at the cost of a reduced Purcell factor ($F_P=6.1$ compared to $F_P=8.5$ exactly at resonance). Consequently, the readout speed is slightly reduced compared to 0\,T. However, we still achieve high-fidelity single-shot spin readout within 3\,ns.

In order to estimate the spin-readout fidelity, we perform Monte Carlo simulations of the single-shot traces with parameters matching our experiment. The simulations include only a few parameters: the overall system efficiency $\eta$, the Purcell factor $F_P$, and the spin-flip time, i.e.\ the relaxation time, $T_1$. At $B=0$, $\eta=37$\%. At B=2.0 T, technical issues result in a slightly reduced efficiency, $\eta=25$\% (Supplementary Sec.\,VI). The spin $T_1=158$\,ns was measured via the quantum jump experiments discussed in the next section. We define the readout-time-dependent fidelity as \cite{Vamivakas2010}
\begin{equation}
   \mathcal{F}(t) = 1 - p_{\textrm{bright}}\cdot e_{\textrm{bright}}(t) -p_{\textrm{dark}}\cdot e_{\textrm{dark}}(t),
   \label{eq:fidelity}
\end{equation}
where $p_{\textrm{bright}}$ ($p_{\textrm{dark}}$) is the occupation probability of the bright (dark) state, and $e_{\textrm{bright}}$ ($e_{\textrm{dark}}$) the respective time-dependent probability of assigning the spin state incorrectly. The spin occupation probability distribution depends on the spin-flip rates, as well as the readout pulse duration and repetition rate; for our experiments it is approximately 50:50 ($\ket{\uparrow}$:$\ket{\downarrow}$). The error $e_{\textrm{bright}}$ is determined on these timescales by imperfect overall system efficiency (which can lead to a spin projected into the bright state being incorrectly assigned as the dark state should no photon be detected). The error $e_{\textrm{dark}}$ is determined by laser leakage (which can lead to a spin projected into the dark state being incorrectly assigned as the bright state). Errors due to spin-flips during the readout time (either due to laser back-action or spin relaxation) play a minor role in our experiment. Our Monte Carlo simulations capture all of these error sources quantitatively ($e_{\textrm{dark}}$\,=\,2.6\%, $e_{\textrm{bright}}$\,=\,6.9\% at 3\,ns; full details of the fidelity calculation and the influence of readout errors can be found in Supplementary Sec.\,VI). The simulated count fractions show very good agreement with our experimental results and allow us to extract a maximum readout fidelity of ($95.2 \pm 0.7$)\% in 3\,ns. The calculated readout fidelity as a function of readout-time for the configurations with $\ket{\uparrow}$ and with $\ket{\downarrow}$ as the bright state is plotted in Figs.\,\ref{fig:Fig2}(c) and (d), respectively.

\begin{figure*}[ht!]
	\centering
	\includegraphics[width=1\textwidth]{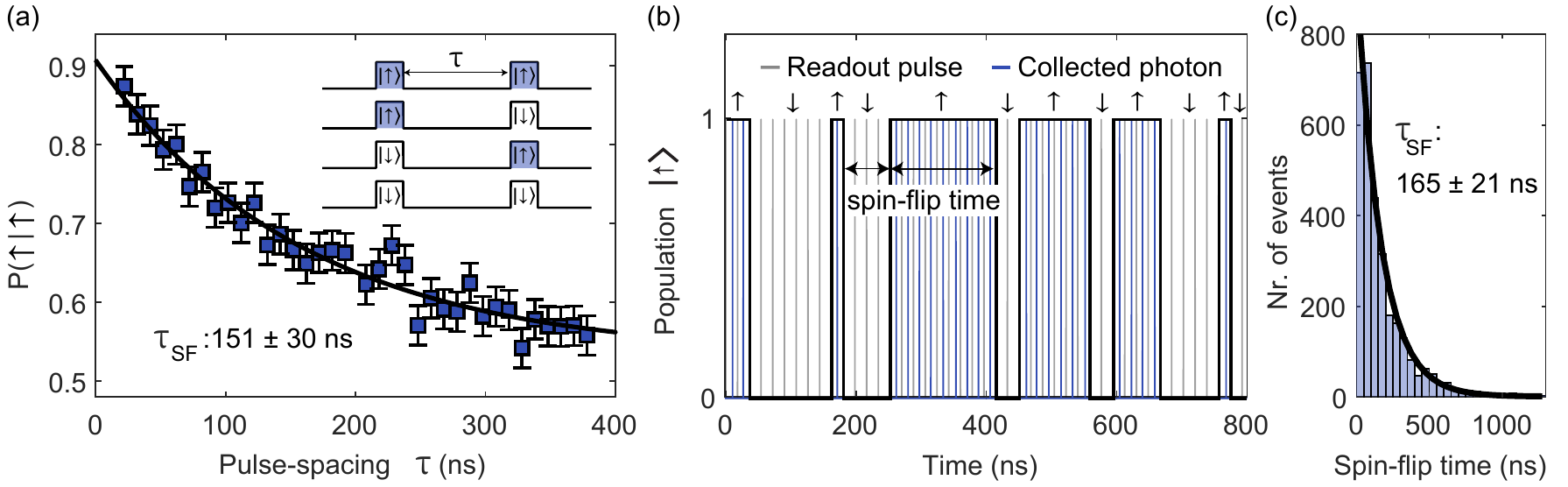}
	\caption{\textbf{Repeated single-shot measurements and quantum jumps.} \textbf{(a)} Conditional probability of measuring $\ket{\uparrow}$ given that the first measurement returned $\ket{\uparrow}$ for two sequential (3\,ns duration) readout pulses as a function of $\tau$, the delay between the pulses. For short values of $\tau$ the second measurement outcome is correlated with the first;	for longer values of $\tau$, the probability to acquire the same outcome decreases exponentially, revealing a spin-flip rate of about 151\,ns. \textbf{(b)} By repeatedly measuring the spin state with excitation pulses spaced by 12\,ns (the detector's dead time), we observe quantum jumps of the spin state. \textbf{(c)} The time between the spin flip events in (b) are extracted for a total measurement time of 2.4\,ms and are summarised in a histogram. The distribution of the events reveals a spin-flip time of 165\,ns, matching well the result of the two-pulse measurements in (a).}
	\label{fig:Fig3}
\end{figure*}
\subsection{Repeated readout and quantum jumps}

The fast spin readout enables us to probe the electron spin dynamics. By repeated single-shot measurements of the spin state, we can determine the spin-flip time from the correlation between sequential measurements. Additionally, we can track the electron spin state in real time, observing quantum jumps as the spin flips. 
In Fig.\,\ref{fig:Fig3}\,(a) we perform a pulse sequence consisting of two readout pulses separated by a time $\tau$. Here we fix the length of both readout pulses to be 3\,ns, and the pulse repetition time to be 400\,ns. The first readout pulse is a projective measurement of the spin state: in effect, the spin is initialised at $\tau=0$ with a fidelity given by either $e_{\textrm{bright}}$ or $e_{\textrm{dark}}$. The second readout pulse can then be used to determine the spin state at $\tau>0$ allowing us to measure the correlation between the two measurement outcomes as a function of $\tau$. Figure\,\ref{fig:Fig3}\,(a) shows the conditional probability of measuring spin $\ket{\uparrow}$ in the second pulse (as a function of $\tau$), given that the first read result returned $\ket{\uparrow}$. We note that the minimum spacing between the two pulses is limited to $\tau\gtrsim12$\,ns by the dead time of the detector. Increasing $\tau$ decreases the probability of reading out the same spin state for both pulses due to spin flips, and for large $\tau$ the second read is completely uncorrelated with the first. By fitting an exponential decay to the data in Fig.\,\ref{fig:Fig3}\,(a), we extract a spin-flip time of $150 \pm 30$\,ns. Furthermore, the limit as $\tau \rightarrow 0$ of this conditional probability is approximately $1-e_{\textrm{bright}}$, confirming the value of $e_{\textrm{bright}}$ determined from the Monte Carlo simulations. Similarly, a measurement of the dark-dark conditional probability confirms the value of $e_{\textrm{dark}}$.

Given that our readout sequence is much shorter than the spin lifetime, we can use repeated single-shot measurements to detect real-time quantum jumps of the electron spin state. For that purpose, we send in a train of 3\,ns readout pulses spaced by the minimum 12\,ns allowed by the detector's dead time.
We observe quantum jumps in the spin state, as shown in Fig.\,\ref{fig:Fig3}(b). (In the original quantum jump experiment, the quantum jumps between the bright and dark states were driven with weak coherent excitation \cite{Bergquist1986}. Here, the jumps are driven by a dissipative process, energy exchange with the Fermi sea via co-tunneling.) The time between spin-flip events during a 2.4\,ms total acquisition period is extracted and summarised in the histogram in Fig.\,\ref{fig:Fig3}(c). From the exponential decay in the number of events per flip time, we can extract the spin-flip time to be approximately 165\,ns, consistent with the results from the double-pulse experiment in Fig.\,\ref{fig:Fig3}(a).

\section{Discussion and Outlook}

We have demonstrated that the frequency-selective Purcell enhancement provided by our optical microcavity enables us to perform single-shot readout of a QD spin state within a few nanoseconds, with a fidelity as high as 95\%. Our results bring the spin readout time for semiconductor QDs close to the short optical spin manipulation times \cite{Press2008, Bodey2019}, and well below previously demonstrated relaxation ($T_1$) \cite{Dreiser2008} and dephasing ($T_{2}^{*}$) times \cite{Gangloff2019,Bodey2019,Jackson2022}. For recent loophole-free Bell tests, entangled nitrogen vacancy (NV) centres were positioned 1.28\,km apart to allow 4.27\,$\mu$s for the Bell sequence to be performed such that the NVs are space-like separated \cite{Hensen2015}. Of this 4.27\,$\mu$s, 3.7\,$\mu$s was used for the single-shot spin readout. Our rapid spin readout indicates that similar Bell tests could be performed using semiconductor QDs located significantly closer together, mitigating the challenge of synchronising experiments between different buildings. By combining the highly indistinguishable photons created by remote semiconductor QDs \cite{Zhai2022}, the high system efficiency of our microcavity \cite{Tomm2021}, along with $T_{2}^{*}$-enhancement via cooling of the nuclear spins \cite{Gangloff2019,Jackson2022}, high-fidelity spin-spin entanglement generation rates of a few hundred MHz are feasible.

We can foresee several ways to improve the readout time in our experiment even further. Most simply, the overall system efficiency can be increased by improving the detector system (fibre couplers and detector itself). Furthermore, it should be possible to operate at the true cavity resonance in an applied magnetic field, thereby at maximum Purcell factor. Our Monte-Carlo simulations show that these changes would allow single-shot readout with a fidelity of 99.5\% in less than one nanosecond to be achieved.

Given the efficient generation of single photons, fast spin initialisation and rotation, and now fast single-shot spin readout, the next step is the implementation of coherent manipulation of the spin-state together with spin readout. Fast spin manipulation relies on a Raman transition that is naturally established in an in-plane magnetic field (Voigt configuration). In this case, the four transitions (Fig.\,\ref{fig:Fig2}(b)) have equal optical dipole moments such that readout back-action is maximal: spin readout becomes challenging. With our approach, this longstanding problem, spin readout in the Voigt geometry, can be solved: the resonant cavity restores a spin-conserving process, i.e.\ a cycling transition; the high overall system efficiency enables a readout outcome before back-action occurs. With our present Purcell enhancement, our simulations show that in the Voigt configuration, readout on the same timescale with a fidelity of up to 89.9\% is feasible. This will enable the spin to be projected out of spin-photon entangled states, a crucial step towards the generation of photonic cluster states.

\section{Acknowledgments}
We acknowledge financial support from Horizon-2020 FET-Open Project QLUSTER, Swiss National Science Foundation project 200020\_204069, and NCCR QSIT. A.J.\ acknowledges support from the European Unions Horizon 2020 Research and Innovation Programme under Marie Sk\l{}odowska-Curie grant agreement no.\ 840453 (HiFig). S.R.V., R.S., A.L.\ and A.D.W.\ gratefully acknowledge support from DFH/UFA CDFA05-06, DFG TRR160, DFG project 383065199 and BMBF Q.Link.X.

\section{Methods}

\subsection{Device and cavity structure}

The semiconductor heterostructure containing the QDs is an n-i-p diode structure grown by molecular beam epitaxy. The tunnel barrier separating the QDs from the back contact is 25 nm-thick. Spin relaxation is dominated by the co-tunneling between a QD electron and the electrons at the Fermi energy in the back contact \cite{Smith2005, Dreiser2008} and results in the relatively short $T_1$-times measured here (see Supplementary Sec.\,I for details). We note that the co-tunneling rate can be decreased exponentially by increasing the thickness of the tunnel barrier and very long $T_1$-times (approaching a second) have been recorded at the magnetic fields used here \cite{Dreiser2008}. The short $T_1$ times used here were useful in that co-tunneling randomises the nuclear spins thereby avoiding optical dragging \cite{Latta2009}. The highly reflective bottom mirror consists of 46 $\lambda/4$-layer pairs of AlAs/GaAs, with a nominal centre wavelength of 940\,nm. Further details of the device design have been previously published \cite{Najer2019,Tomm2021}.

The top mirror of the tunable microcavity is fabricated using CO$_2$-laser ablation of a fused silica substrate to create a concave mirror shape with radius of curvature $R=12$\,$\mu$m. After laser ablation, the top mirror is coated with 8 $\lambda/4$-layer pairs of Ta$_2$O$_5$ ($n=2.09$ at $\lambda=920$\,nm) and SiO$_2$ ($n=1.48$ at $\lambda=920$\,nm). The number of layer pairs was designed to maximise the photon extraction efficiency from the system \cite{Tomm2021}. The cavity exhibits a mode splitting between orthogonal horizontal (H) and vertical (V) polarised modes due to a slight birefringence in the semiconductor sample (50\,GHz in our experiments). Each cavity mode has a Q-factor of 16,000.

\vspace{5mm}

\subsection{Optical measurement setup}

We use a cross-polarised microscope setup to separate the resonant excitation laser light from the QD emission \cite{Kuhlmann2013_darkfield}. The mode splitting between the orthogonally polarised modes of our microcavity allows us to excite the QD using the Lorentzian ``tail'' of one mode and to collect via the other mode \cite{Wang2019, Tomm2021}. The QD photons emitted from the cavity are then elliptically polarised, with the major polarisation axis orthogonal to the excitation laser polarisation. Using this technique, we overcome what would otherwise be a 50\% loss of QD photons in the cross-polarisation optics \cite{Wang2019}.

At 0.0\,T we measure a Purcell factor of $F_P=8.5$ (see Supplementary Sec.\,IV), where $F_P$ is the enhancement of the emitter's decay rate in the cavity compared to the decay rate in the bare sample. At 2.0\,T we observe poor laser suppression in the cross-polarised microscope head at the exact cavity resonance, which we attribute to a Faraday effect in the objective lens and/or cavity top mirror. However, the laser suppression is recovered at a frequency detuned by 7.5\,GHz from the exact cavity resonance (see Supplementary Sec.\,IV for further details). For spin readout we operate at optimal laser suppression where the Purcell factor is reduced to $F_P=6.1$.

We generate short (1$-$10\,ns) optical readout pulses using a high-bandwidth EOM (EOSpace AZ-6S5-10-PFA-SFAP-950-R5-UL) driven by a high sampling-rate AWG (Tektronix 7122C). Photons emitted from the QD are detected using a superconducting nanowire single photon detector (SingleQuantum, efficiency 82\% at 920\,nm), and detection events are registered using a time-correlated single photon counting module (Swabian Instruments Timetagger Ultra).

The full cavity stack is mounted in a liquid helium cryostat operating at 4.2\,K, equipped with a superconducting solenoid magnet. We operate at 2.0\,T for the spin readout experiments.

\newpage
\onecolumngrid
\section*{Supplementary Information: Cavity-enhanced single-shot readout of a quantum dot spin within 3\,nanoseconds}
\setcounter{section}{0}
	
\section{Extraction of spin lifetime from $\mathbf{g^{(2)}(\tau)}$}
\label{sec:g2s}
We can characterise the spin-flip rate in our experiment by measuring the second-order correlation function of the resonance fluorescence $g^{(2)}(\tau)$. By driving only one of the Zeeman-split trion states, spin-flips are observed as blinking in the quantum dot (QD) fluorescence as the spin state switches between the on-resonance (bright) and off-resonance (dark) states. This blinking results in bunching of the $g^{(2)}(\tau)$-function, which for spin-flip processes occurring with an exponential distribution in time, is described by the expression \cite{Sychugov2005}:
\begin{equation}
    g^{(2)}(\tau) = 1+\frac{\tau_\mathrm{off}}{\tau_\mathrm{on}}\exp\left[-\left(\frac{1}{\tau_\mathrm{on}}+\frac{1}{\tau_\mathrm{off}}\right)\tau\right],
\label{eq:g2_bunching_expression}
\end{equation}
where $\tau_\mathrm{on}$ is the average time the spin remains in the bright state before flipping, and $\tau_\mathrm{off}$ is the average time the spin remains the the dark state.

\begin{figure}[b!]
	\centering
	\includegraphics{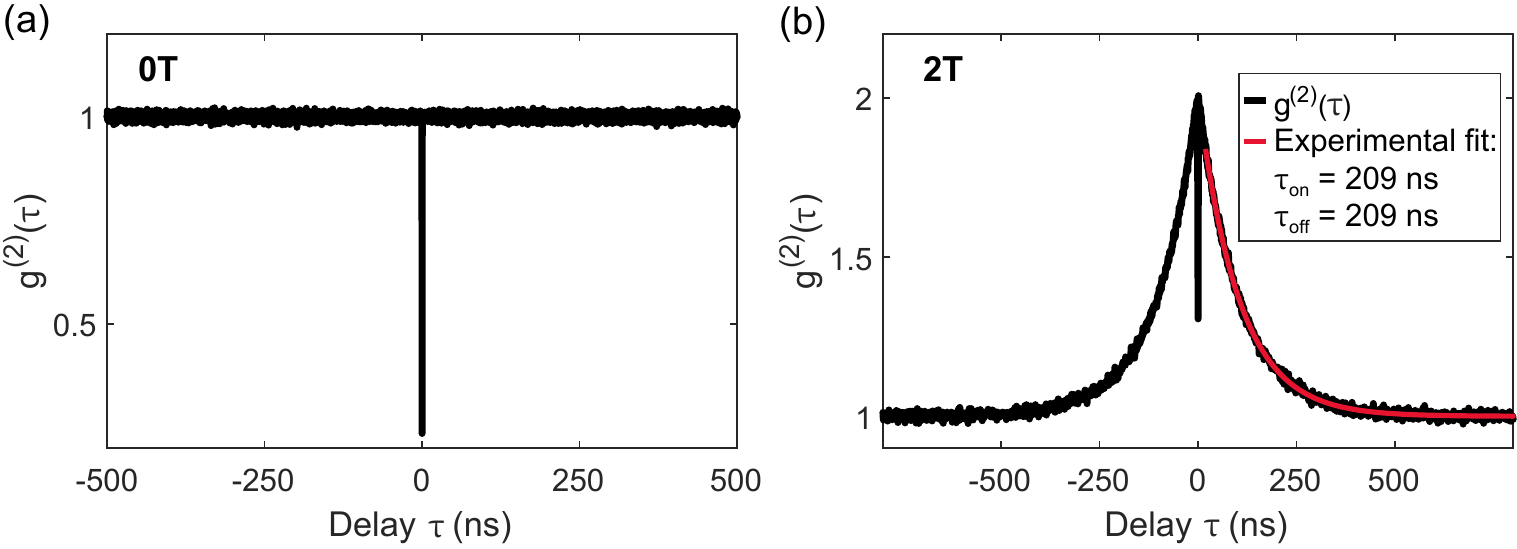}
	\caption{\textbf{Second-order correlation function at 0\,T and 2\,T.}
	\textbf{(a)} At $B=0$\,T the $g^{(2)}(\tau)$ features anti-bunching at $\tau=0$ but no bunching out to $\tau=500$\,ns, demonstrating that the QD emission is stable on this timescale.
	\textbf{(b)} $g^{(2)}(\tau)$ at $B=2$\,T while resonantly driving the higher-frequency trion state. The QD emission shows a clear bunching. We fit the $g^{(2)}(\tau)$ with an exponential decay to determine the spin-flip rate. The data shown here is acquired with a laser power equal to the QD saturation power; from the fit we extract $\tau_\mathrm{on}=\tau_\mathrm{off}=209$\,ns.
	}
	\label{fig:bunching_vs_mag_field}
\end{figure}

This blinking is not present at zero magnetic field, where the trion states are degenerate such that both spin states are driven with a resonant linearly-polarised laser. Figure \ref{fig:bunching_vs_mag_field}(a) shows $g^{(2)}(\tau)$ measured at zero magnetic field for the X$^-$ transition of the QD used in our experiments. As expected for a single emitter, anti-bunching is observed at $\tau=0$. Away from $\tau=0$, the $g^{(2)}(\tau)$ is flat with no significant bunching observed, indicating that the QD emission is stable.
Figure \ref{fig:bunching_vs_mag_field}\,(b) shows a similar $g^{(2)}(\tau)$-measurement, now for an out-of-plane magnetic field $B=2.0$\,T. Here, the laser drives the higher-frequency trion state, and clear bunching is observed. By fitting the measured $g^{(2)}(\tau)$ to Eq.\,\ref{eq:g2_bunching_expression}, we can extract the characteristic timescale on which the QD emission switches on and off. Because the switching between bright and dark states occurs purely due to spin flips, the timescale of the bunching decay is a direct measurement of the spin lifetime.

\begin{figure}[t]
	\centering
	\includegraphics{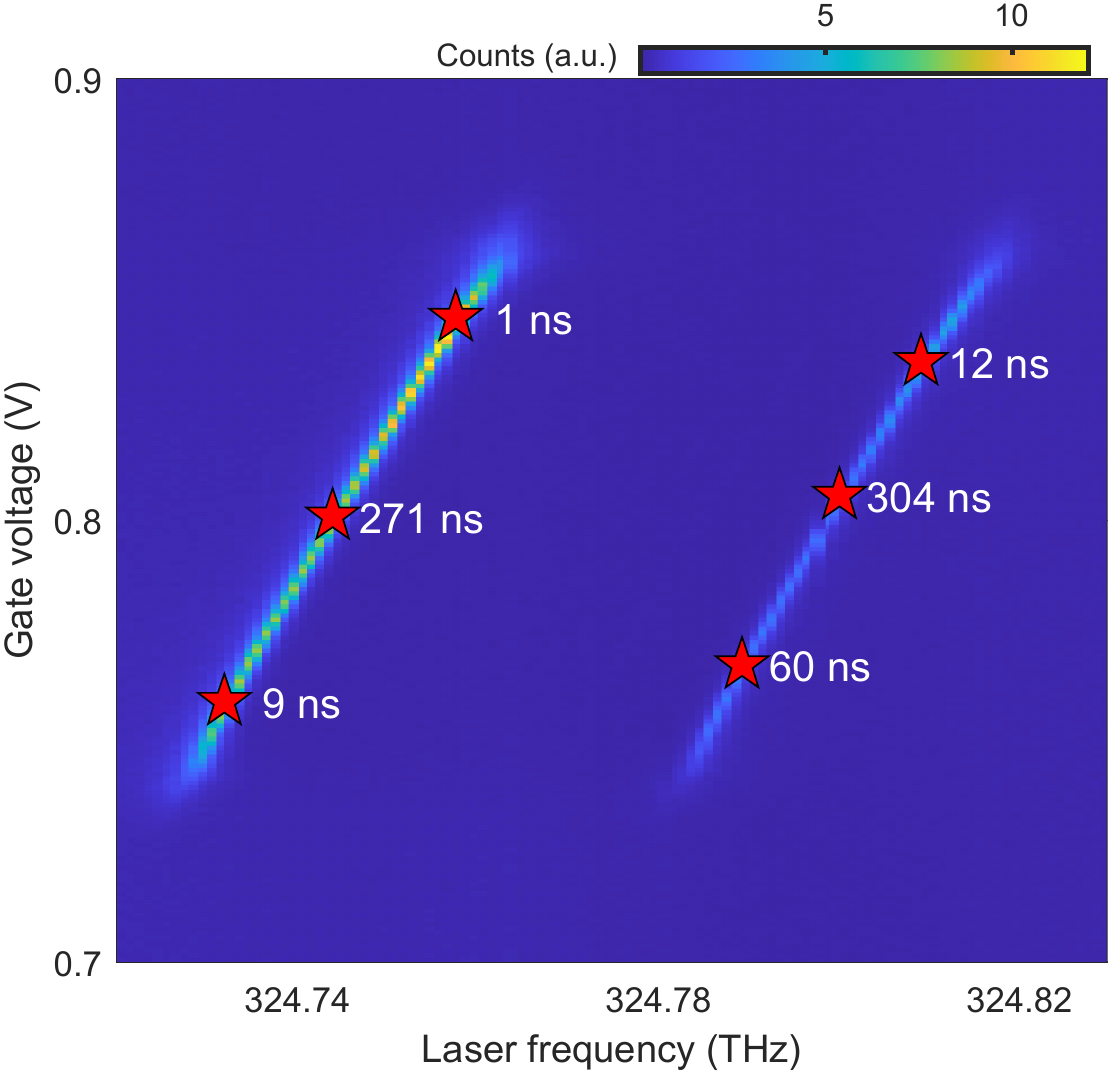}
	\caption{\textbf{Spin lifetime as a function of position within the charge plateau.} To map out the charge plateaus, the QD fluorescence was collected as a function of excitation laser frequency and gate voltage applied across the diode structure. The applied magnetic field is 2.0\,T, resulting in a splitting of $\sim55$\,GHz between the two vertical optical transitions. The red stars indicate positions on the Zeeman-split plateaus for which the spin lifetime was measured (as described in Fig.\,\ref{fig:bunching_vs_mag_field}(b)). The spin lifetime decreases at the plateau edges to very small values, a clear sign of co-tunnelling. However, even in the plateau centre the longest spin lifetime time we observe is $\sim$300\,ns, also determined by co-tunnelling. The spin lifetime numbers correspond to $\tau_\mathrm{on}$, which is approximately equal to $\tau_\mathrm{off}$ for the low laser powers (well below saturation) used in this measurement. We note that to acquire this data we adjust the cavity length when the laser frequency is stepped such that the laser remains on resonance with the cavity.
	}
	\label{fig:plateau_lifetime_map}
\end{figure}

We measured the $g^{(2)}(\tau)$ and extracted $\tau_\mathrm{on}$ and $\tau_\mathrm{off}$ for several different positions on the X$^-$ charge plateau. The results are summarised in Fig.\,\ref{fig:plateau_lifetime_map}, where the red stars indicate the charge plateau position at which each experiment was performed. Here the numbers on the plot correspond to $\tau_\mathrm{on}$. The data in Fig.\,\ref{fig:plateau_lifetime_map} were acquired using low laser powers, significantly below the saturation power. The spin lifetime is very short, a few nanoseconds, at the edges of the charging plateau, and reaches a modest value, $\sim$\,300\,ns, at the centre of the plateau. These are the hallmarks of co-tunneling \cite{Smith2005,Dreiser2008}, a process in which a combined tunnelling process swaps an electron confined to the QD with an electron close to the Fermi energy in the Fermi sea. From the measured spin lifetimes, it is clear that co-tunneling determines the spin lifetime even at the centre of the plateau. The observed maximum spin lifetime of 300\,ns is orders-of-magnitude less than the expected intrinsic spin lifetime via a phonon-mediated process at this magnetic field: previous experiments using InGaAs QDs have demonstrated $\sim20$\,ms at similar magnetic field strengths \cite{Dreiser2008,Lu2010}. The relatively fast co-tunneling is a consequence of the 25 nm-thick tunnel barrier, the distance separating the back contact and QD-layer in the heterostructure. 

We note that due to the high speed at which we can perform single-shot spin readout, the relatively short co-tunneling induced spin-flip time that we observe is not the limiting factor for the readout fidelity. For a readout time of 3\,ns, we would expect a spin flip during the readout pulse in only $1-\exp(-3/300)\sim1$\% of readout attempts.

For the low powers used in Fig.\,\ref{fig:plateau_lifetime_map}, we do not observe spin pumping, for which the typical signature is a region of decreased signal at the centre of the charge plateau: in spin pumping, the excitation results in occupation of the dark state \cite{Atature2006,javadi_spinphoton_2018}. The absence of a spin pumping signature in Fig.\,\ref{fig:plateau_lifetime_map} indicates that the spin pumping rate is significantly smaller than the spin flip rate. Spin pumping arises via spin-nonconserving spontaneous emission, a ``diagonal" transition, Fig.\, 2(b). (The spin-conserving recombination is the ``vertical" transition, Fig.\,2(b).) The branching ratio is the ratio of the diagonal to vertical recombination times. It can be inferred from the $g^{(2)}(\tau)$ recorded with optical driving powers above saturation power. Specifically, the branching ratio can be extracted from $g^{(2)}(\tau)$ by solving the incoherent part of the optical Bloch equations, i.e.\ the rate equations describing the populations of the three relevant QD levels \cite{Gaebel2004}. Following this process, we extract a branching ratio of $\Gamma_s/\gamma_d=600\pm200$ (where $\Gamma_s$ is the vertical spin-conserving decay rate, and $\gamma_d$ the diagonal spin-nonconserving rate). This branching ratio applies to the experimental conditions for the readout process in which one of the vertical transitions is in resonance with the cavity.

Spin-nonconserving spontaneous emission is the origin of back-action in the spin readout process. For single-shot readout, the branching ratio must be high enough for the spin state to be assigned with high fidelity before a laser-induced spin-flip transition occurs. This is the case here. In fact, the branching ratio is sufficiently high that back-action is a negligible source of readout error in these experiments.

\clearpage
\section{Spin initialisation by optical pumping}
\label{sec:initialisation}
\begin{figure}[b]
	\centering
	\includegraphics{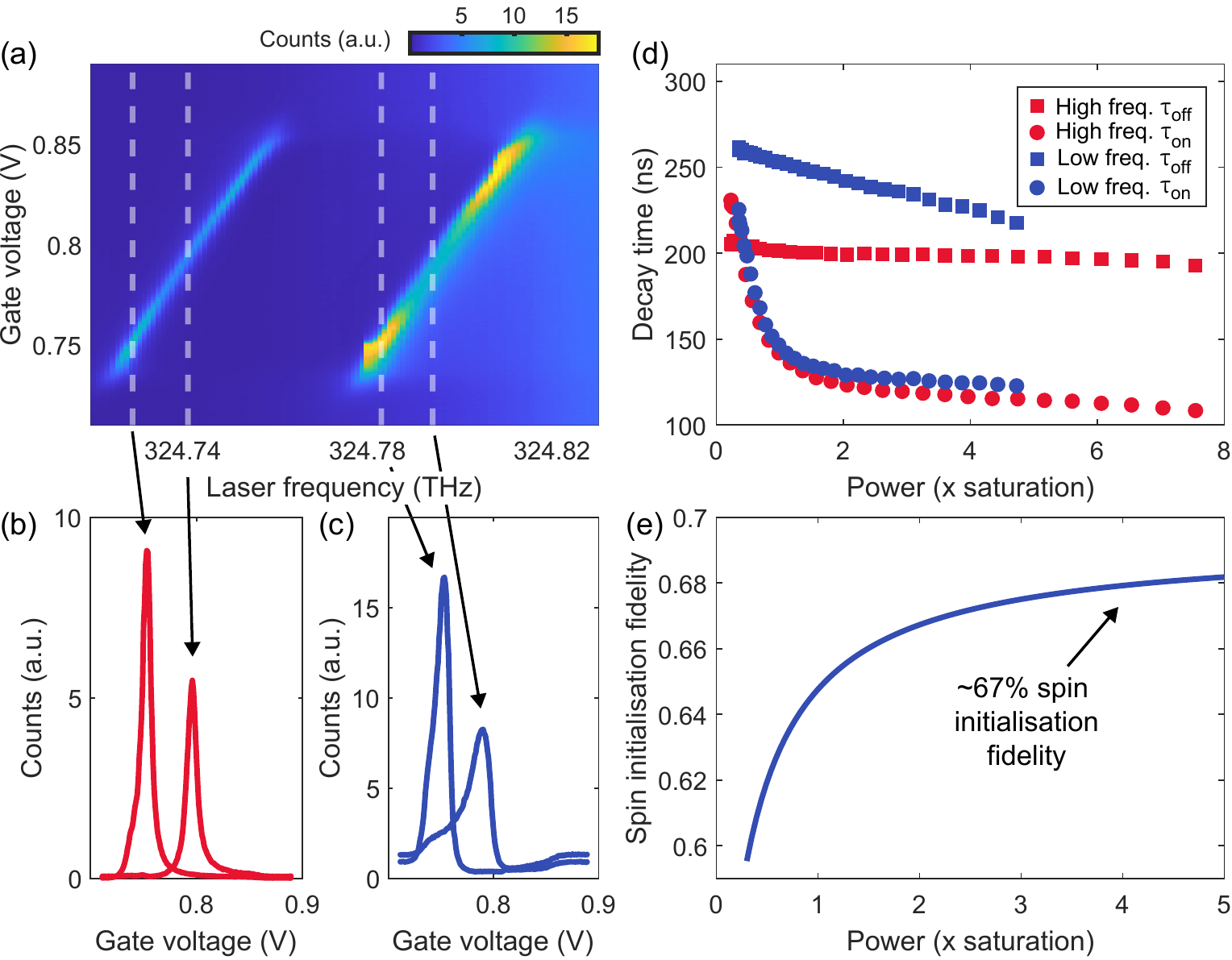}
	\caption{\textbf{Partial optical spin initialisation.} 
	\textbf{(a)} Plateau map acquired with a laser power twice the saturation power. We observe a reduction in signal at the plateau centre, consistent with optical spin pumping. However, comparing the signal in the centre of plateau with that at the edges in the fast co-tunnelling regime suggests a poor spin initialisation fidelity.
	\textbf{(b)} Comparison of fluorescence signal for the low-frequency trion in the fast co-tunnelling regime (left, lower gate voltage) and in the plateau centre (right, higher gate voltage).
	\textbf{(c)} Similar to (b) but for the high-frequency trion.
	\textbf{(d)} Spin-flip times $\tau_\mathrm{on}$ and $\tau_\mathrm{off}$ as a function of excitation laser power measured at the centre of the charge plateau, resonantly driving the low-frequency trion (blue data points) or the high-frequency trion (red data points).
	\textbf{(e)} Estimated spin initialisation fidelity from solving the incoherent part of three-level optical Bloch equations, using spin-flip rates extracted from (d). Based on this analysis, a maximum spin initialisation fidelity of $\sim 67$\% is expected.
	}
	\label{fig:optical_pumping}
\end{figure}

One strategy for demonstrating single-shot spin readout is to first initialise the spin in a known state, then perform the readout sequence. By comparing the spin state attributed during readout with the initially prepared state, the readout fidelity can be quantified. This method relies on the ability to initialise the spin state with high fidelity. In our experiments, the combination of a modest spin lifetime together with a large branching ratio make the initialisation of a known spin state via optical pumping challenging. The maximum spin pumping rate that we were able to achieve was comparable to the co-tunnelling rate in the centre of the plateau. Figure\,\ref{fig:optical_pumping}(a) shows a plateau map similar to Fig.\,\ref{fig:plateau_lifetime_map} using a significantly higher excitation laser power (approximately twice the saturation power). We now observe decreased fluorescence intensity in the plateau centre, consistent with spin pumping. However, rather than near-complete extinction of the fluorescence, the signal at the plateau centre is reduced by only a factor of about two (Figs.\,\ref{fig:optical_pumping}(b),(c)) compared to at the edges where rapid co-tunnelling prohibits spin pumping. The incomplete suppression of fluorescence in the centre of the charge plateau indicates that our spin initialisation fidelity using optical spin pumping is likely to be modest.

To quantify the achievable spin initialisation fidelity, we measured the spin-flip rates $\tau_\mathrm{on}$ and $\tau_\mathrm{off}$ (as shown in Fig.\,\ref{fig:optical_pumping}(d)) near the centre of the charge plateau as a function of laser power. The result is shown in Fig.\,\ref{fig:optical_pumping}\,(d); the four curves show $\tau_\mathrm{on}$ and $\tau_\mathrm{off}$ for the laser resonant with the high-frequency trion transition or the lower-frequency trion transition.

We solve the incoherent part of the optical Bloch equations to estimate the population of the spin initialisation target state (equivalent to the initialisation fidelity) as a function of the initialisation laser power, shown in Fig.\,\ref{fig:optical_pumping}\,(e). We find that the achievable initialisation fidelity saturates to a rather low value; for a laser power of 4x the QD saturation power, the initialisation fidelity (defined as $|\langle \psi_{\rm actual}|\psi_{\rm target}\rangle|^2$) is approximately 67\%.

With such a low initialisation fidelity, a measurement sequence of first initialising the spin before readout is impractical, as the initialisation fidelity would dominate the total sequence fidelity and obscure the actual readout error. Instead, to characterise our single-shot readout fidelity we repeat our readout sequence with a delay comparable to the plateau-centre spin lifetime, which results in an approximately 50:50 spin state occupation probability over the course of a large number of sequential readout sequences. As discussed in Sec.\,\ref{sec:readout_simulation}, by characterising each readout error process individually we can determine the overall readout fidelity.

We stress that the inability to initialise the spin with high fidelity in these experiments is a consequence of the tunnel barrier thickness and does not represent a limitation of the scheme itself. High initialisation fidelities can be achieved by suppressing the co-tunneling at the plateau centre using a larger tunnel barrier \cite{Atature2006}. 

\newpage
\section{Continuous wave quantum jumps measurements}
\label{sec:CWjumps}
\begin{figure}[h]
	\centering
	\includegraphics{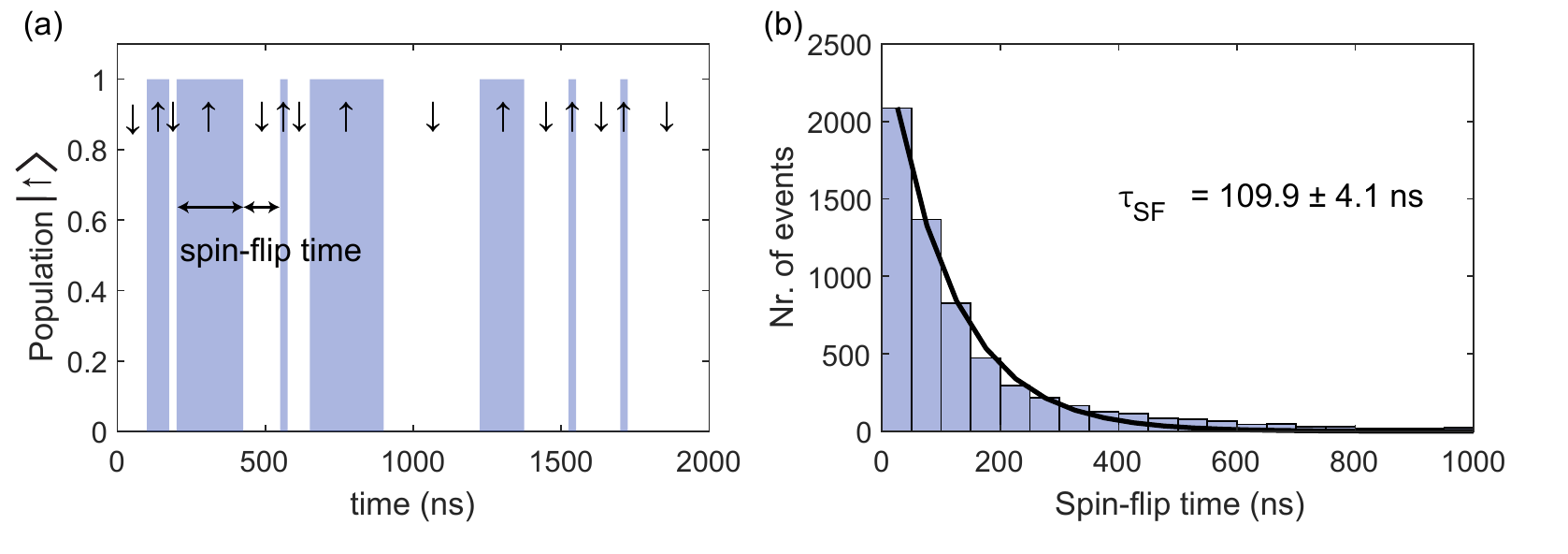}
	\caption{\textbf{CW Quantum Jumps (a)} Normalised photon counts as a function of detection time. The quantum jumps between the two spin states can be observed and the time over which the spin remains the same can be extracted. \textbf{(b)} Histogram of the extracted times between spin-flip events reveals an exponential decay. The spin-flip time is extracted from an exponential fit to be (109.9$\pm$4.1)\,ns.}
	\label{fig:cw_qjumps}
\end{figure}

Complementary to the experiments demonstrating quantum jumps using rapidly repeated readout pulses shown in Fig.\,3(b) of the main text, we also observed quantum jumps using continuous wave (CW) excitation. A CW laser set to four times the saturation power of the bright state transition was used, and the emitted photons were routed (via cascaded 50:50 beam splitters) to four SNSPD detectors. In contrast to the pulsed single-shot readout experiments in the main text (where only one SNSPD detector was used), we used four detectors to mitigate partially the impact of the detectors' dead time. We note that the addition of the cascaded beam splitters reduces the overall system efficiency, hence why we focused on the pulsed quantum jumps experiments in the main text. 
We measured the signal on all four detectors simultaneously, and the resulting counts registered by the four detectors were then added together. If at least one photon is measured in a time-bin, the state is assigned spin up ($\ket{\uparrow}$). If no photon is detected, the state is assigned spin down ($\ket{\downarrow}$). A fraction of these quantum jumps is shown in Fig.\,\ref{fig:cw_qjumps}(a). 

The CW quantum jumps we observe provide an additional method to characterise the spin-flip rates in our system, as $\tau_\mathrm{on}$ and $\tau_\mathrm{off}$ can be directly extracted from the waiting-time distributions for $\ket{\uparrow}$ and $\ket{\downarrow}$.
The time over which the spin state remains the same is extracted over an experiment of 50\,ms duration, and its distribution is shown as a histogram in Fig.\,\ref{fig:cw_qjumps}(b). By fitting the decay in the histogram we determine a spin-flip time of $\tau_\mathrm{SF}=$ 109.9.9$\pm$4.1\,ns (where $\tau_\mathrm{SF}$ is the average of $\tau_\mathrm{on}$ and $\tau_\mathrm{off}$). The result is slightly lower (although broadly consistent) with the spin-flip time extracted from the $g^{(2)}(\tau)$ recorded using the same laser power (Sec.\,\ref{sec:g2s}); the present experiment was performed slightly offset from the exact charge plateau centre, which may explain the difference. Due to partial spin-pumping with CW excitation (see Sec.\,\ref{sec:initialisation}), the observed spin-flip time is slightly smaller than that measured with pulsed excitation (165\,ns, main text).

\newpage
\section{Cavity suppression at 2.0\,T}

\label{sec:cavitysuppr}

\begin{figure}[h]
	\centering
	\includegraphics{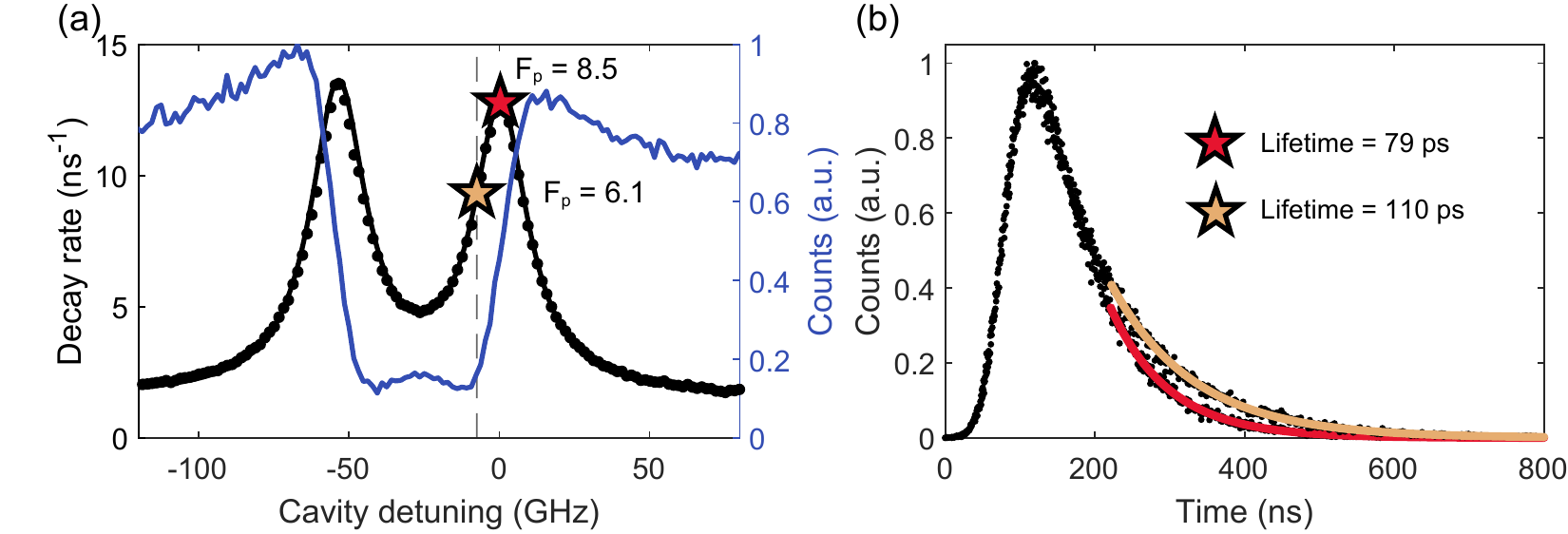}
	\caption{\textbf{(a)} Laser leakage into the collection channel (blue) and decay rate of the QD (black) as a function of cavity detuning. The laser leakage shows two minima and neither aligns perfectly with the cavity resonance. At the cavity resonance, the Purcell factor is $F_P=8.5$; at the lowest laser leakage, $F_P=6.1$. \textbf{(b)} Time-resolved lifetime measurement on resonance with the cavity (red) and at the detuning for minimum laser leakage (yellow). The response of the QD to a short excitation pulse (few ps) is measured and reveals an exponential decay (black). The lifetime is extracted from an exponential fit and is 79\,ps at resonance and 110\,ps at minimum laser leakage.} 
	\label{fig:cavity}
\end{figure}

In order to readout the spin-state correctly, the excitation laser has to be suppressed well enough to minimise spurious counts on the detector due to laser leakage. Otherwise, there is a significant probability that the readout pulse projects the spin into the dark spin-state yet the outcome is recorded falsely as the bright spin-state. A cross-polarisation setup is used to prevent laser light from entering the detection fibre \cite{Kuhlmann2013_darkfield}. This works extremely well at zero magnetic field. However, in an applied magnetic field, the background suppression works slightly less well. This effect likely arises from a Faraday effect in the top mirror of the cavity and/or objective lens. A normalised background signal showing the counts due to laser leakage as a function of cavity detuning is shown in Fig.\,\ref{fig:cavity}(a). Unfortunately, the point of maximum laser suppression is not aligned with the cavity resonance, but detuned by 7.5\,GHz. At the cavity resonance, the background is high enough to give a spurious count on the detector in 80\,$\%$ of the readout pulse repetitions, making this regime impractical for the readout. The 2\,T measurements are therefore performed at the cavity detuning where the background is a minimum. At this cavity detuning, the probability of detecting a photon via laser leakage reduces to 1.4\% for a 3\,ns readout pulse.

An important parameter for the spin-readout is the $\beta$-factor which itself depends on the Purcell factor, $F_P$: $\beta = F_P/(F_P +1)$. Cavity-enhanced spin readout depends on achieving $\beta$-factors as close as possible to one, equivalently large Purcell factors. We extract the Purcell-factor as a function of cavity detuning by measuring the lifetime of the QD at each cavity detuning and deriving it via $ \Gamma = F_P \cdot \gamma$, where $\Gamma$ is the Purcell-enhanced decay rate and $\gamma$ is the bare decay rate ($\gamma \approx 0.3$\,GHz). (The decay rate is the inverse of the lifetime, $\gamma = 1/\tau$).  On resonance with the cavity, $F_P = 8.5$. At the detuning for which the laser suppression works best, the Purcell factor is slightly lower, $F_P = 6.1$. The corresponding decay curves following excitation with a few-ps laser pulse are shown in Fig.\,\ref{fig:cavity}(b). The spin read-out experiments were carried out at $F_P = 6.1$.

\newpage
\section{Monte-Carlo Simulation: Count-traces and Fidelity}
\label{sec:readout_simulation}
Our single-shot readout results are modelled using a Monte-Carlo approach in order to determine the readout fidelity. 

\subsection{Simulation of Readout Count fractions}
The simulations of the count fractions are based on a Monte-Carlo method in which the (simulated) readout outcome is recorded many times (100,000 repetitions) in order to mimic the experiment. The readout pulse is considerably longer than the Purcell-enhanced radiative lifetime. The power is also well above the saturation power. These two factors mean that should the spin be projected into the bright state, the exciton population is close to 0.5. The photon emission rate is the occupation of the bright state divided by the lifetime. Each photon is detected with a certain probability, the overall system efficiency. In other words, the detection rate is the emission rate multiplied by the overall system efficiency, $\eta$. A readout cycle is repeated until a photon is detected and the detection time is recorded. Summing up over all repetitions leads to count fractions as in the measurements in Fig.\,2(a) of the main text. The model has four input parameters: the overall system efficiency $\eta$ (the probability that an exciton in the QD results in a click on the detector), the Purcell factor $F_P$, the spin-flip time $\tau_{\textrm{SF}}$, and the probability of detecting a laser photon (to simulate the laser background, see analysis in Sec.\,\ref{sec:cavitysuppr}).

\begin{figure}[b]
	\centering
	\includegraphics[width=\textwidth]{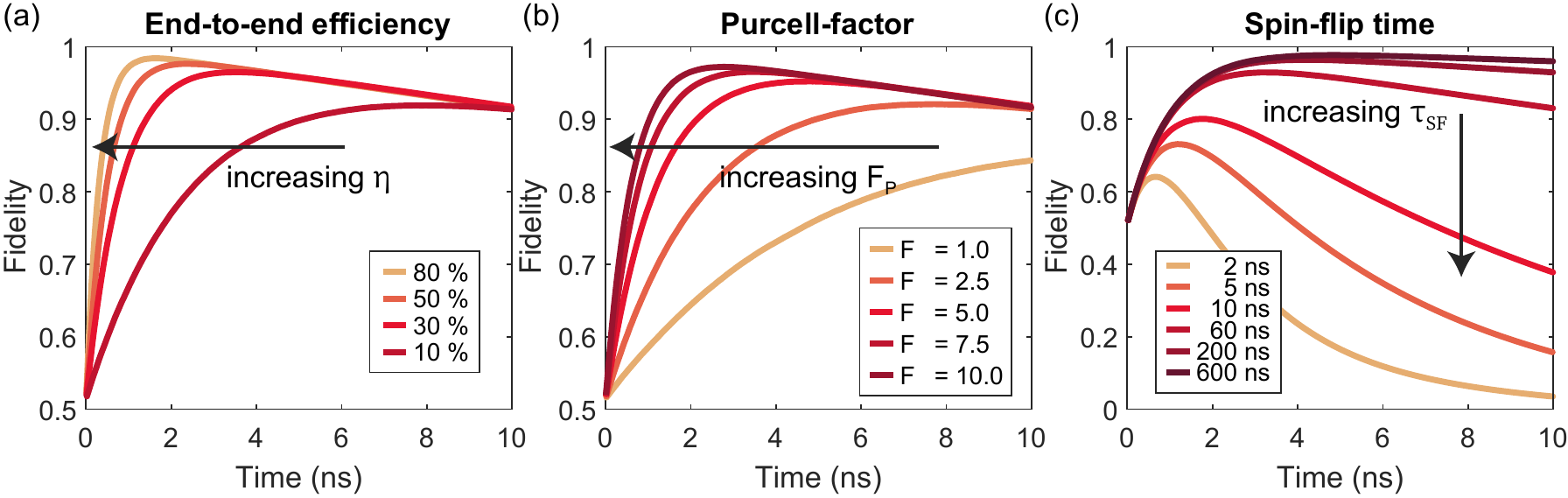}
	\caption{Simulation of the fidelity as a function of readout time for different \textbf{(a)} overall efficiencies, \textbf{(b)} Purcell factors and \textbf{(c)} spin-flip times. While one of the parameters is varied, the other two are set to the experimental conditions: $\eta = 25\%$, $F_P = 6.1$, and $\tau_{\textrm{SF}} = 158$\,ns. All simulations are performed at 6x the saturation power including 1.4\% laser leakage at 3\,ns readout duration.}
	\label{fig:timetraces}
\end{figure}

The dependence of the readout on the overall system efficiency is shown in Fig.\,\ref{fig:timetraces}(a). The higher the efficiency, the sooner the spin-state can be read out, and the lower the probability of incorrectly assigning the spin state. In practice, the overall system efficiency $\eta$ is known based on the analysis in Ref.\,\cite{Tomm2021}, the properties of the fibre couplers, and the quantum efficiency of the detector:
\begin{equation}
    \eta = \beta \cdot \frac{\kappa_{\textrm{top}}}{\kappa+\gamma} \cdot \eta_{\textrm{optics}} \cdot \eta_{\textrm{coupler}}\cdot \eta_{\textrm{detector}}
    \label{eq:efficieny}
\end{equation}
where $\beta$ is the probability that an exciton creates a photon in the H-polarised cavity mode; $\kappa_{\textrm{top}}/(\kappa+\gamma)=96$\% is the probability that a photon in the cavity exits the top mirror; and $\eta_{\rm optics}=69$\% represents the throughput of the optical system from microcavity to the output of the final output fibre (as defined and measured in Ref.\,\cite{Tomm2021}). The output of this fibre is coupled to the detector with an optical coupler (in practice, two fibre-couplers) with efficiency $\eta_{\textrm{coupler}}=80$\%. Finally, the detector has a quantum efficiency of $\eta_{\textrm{detector}}=82$\%.

At $B=0$, $\beta=86$\% such that $\eta=37$\%. We stress that this is the predicted overall system efficiency based on the analysis of all the individual contributions, including the detector efficiency. In practice, this predicted value of $\eta$ describes the experimental results extremely well.

At $B=2$\,T, $\beta=80$\%, resulting in a predicted overall efficiency of $\eta=35$\%. In practice, a slightly lower $\eta$ is required to describe quantitatively the experimental results, $\eta=25$\%. The origin of this slight reduction in $\eta$ with respect to $B=0$ is unknown. For the cavity alignment with $\mathrm{F_p}=6.1$ used in our experiments (shown in Fig. \ref{fig:cavity}\,(a)) the Purcell factor is very sensitive to the exact cavity detuning; a small shift could result in a slightly lower Purcell factor. The effect of a lower Purcell factor in our simulations is similar to that of a lower efficiency. Other explanations could be a deterioration in either the in-coupling efficiency (such that the power exceeds the saturation power by a smaller margin that at $B=0$) or the optical alignment thereby reducing the product $\eta_{\textrm{optics}} \cdot \eta_{\textrm{coupler}}$. We stress that this slight discrepancy between the predicted overall efficiency and the overall efficiency that matches best our experimental data has no impact on our readout fidelity analysis.

\subsection{Calculation of the Readout Fidelity}

The fidelity of the spin-readout is defined as
\begin{equation}
   \mathcal{F}(t) = 1 - p_{\textrm{bright}}\cdot e_{\textrm{bright}}(t) -p_{\textrm{dark}}\cdot e_{\textrm{dark}}(t),
   \label{eq:fidelity}
\end{equation}
where in our experiments both $p_{\textrm{bright}}$ and $p_{\textrm{dark}}$ are approximately equal to 50\%.

The readout projects the spin into either the bright state or the dark state; the readout process records an outcome, either bright or dark. If the spin is projected into the bright (dark) state but readout as dark (bright) then the error is $e_{\textrm{bright}}$ ($e_{\textrm{dark}}$). These error probabilities are time-dependent in that they depend on the duration of the readout pulse.

The errors in the readout have several origins. The bright spin state is correctly assigned if a photon is measured. Therefore, photon loss is an important source of readout error. The overall system efficiency $\eta$ therefore contributes to $e_{\textrm{bright}}$. This source of error can be quantified by isolating this loss process in a simulation which takes the experimental value of $\eta$ but without back-action and without a spin-flip process. This results in $C(t)$, the cumulative distribution function for collecting a count as a function of time induced by a readout pulse starting at $t=0$. In our experiments the error probability $e_{\textrm{bright}}=6.9\%$ at 3\,ns, as shown in Fig.\,2(d) in the main text. Another source of error for the bright state readout is a spin flip during the readout process: the QD can be projected into the bright state by the readout pulse but if it flips to the dark state before a photon is detected the spin is assigned incorrectly. 

These combined contributions to $e_{\textrm{bright}}$ result in: 
\begin{equation}  
    e_{\textrm{bright}}(t) = 1 - C(t) + C(t)\cdot[1-\exp(-t/\tau_{SF})] = 1 - C(t)\cdot \exp(-t/\tau_{SF}),
    \label{eq:bright}
\end{equation}
where $\tau_{SF}$ is the spin-flip time.

The dark state readout error also has two origins. First, the readout can project the spin into the dark state yet be recorded as the bright state should a laser photon leak into the collection channel and be detected. This error can be estimated and taken into account by measuring the count fraction $C_d(t)$ on turning off the QD, i.e.\ detuning the QD with respect to the readout laser (in practice via the gate voltage). Second, as for the bright state, a spin-flip can lead to an error: the readout can project the spin into the dark state yet be recorded as bright if a spin-flip from dark-to-bright state occurs followed by photon detection. The analysis of the second error is more complicated than that of the first. If the spin flips from the dark to the bright state, a photon can be emitted and counted. This takes place with the same time-dependence as $C(t)$, but shifted in time by the location in time of the spin-flip. This effect can be taken into account by a convolution of the shifted count fraction with the spin-flip probability. The combined readout error is therefore:
\begin{equation}
    e_{dark}(t) = C_d(t) + \frac{1}{t}\int_0^t C(t - \tau)\cdot[1-\exp(-t/\tau_{SF})] \cdot d\tau.
    \label{eq:dark}
\end{equation}

For a 3\,ns readout pulse, the error probability $e_{\textrm{dark}}=2.6\%$. For short readout times, the error in reading out the state $e_{\textrm{bright}}$ is high, as not enough time has elapsed to ensure that one of the QD photon is detected by the detector. For longer readout times, the probability of a spin-flip increases, and hence the probability of detecting a photon from the dark state via a spin flip to the bright state ($e_{\textrm{dark}}$) increases. Hence, there is an optimal readout time for which the fidelity can be maximised. By plugging Eq.\,\ref{eq:bright} and \ref{eq:dark} into Eq.\,\ref{eq:fidelity}, we can calculate the fidelity of the readout as a function of the readout time; this is shown in Fig.\,2(c) in the main text. We carry out this calculation on tuning the cavity to the higher-frequency trion and, separately, on tuning the cavity to the lower-frequency trion. The readout fidelity reaches 95.2\% in 3\,ns.

Figure \ref{fig:timetraces} shows the dependence of the fidelity on the end-to-end efficiency (Fig. \ref{fig:timetraces}\,(a)), Purcell factor (Fig. \ref{fig:timetraces}\,(b)) and spin-flip time ((Fig. \ref{fig:timetraces}\,(c)). In Fig. \ref{fig:timetraces} when one of the parameters is varied, the others are set to match our present experimental conditions. However, for realistic improvements to all of these parameters simultaneously the readout can be significantly improved, as we now discuss.

\subsection{Predictions for optimised system and Voigt geometry}
Based on the success of the Monte-Carlo method in describing the experimental results, we can estimate the achievable fidelity for an optimised system as well as for single-shot readout in the Voigt geometry. We assume that the issue of imperfect laser suppression at the exact cavity resonance (Sec.\,\ref{sec:cavitysuppr}) can be overcome. We assume also that a QD can be selected with a larger optical dipole moment -- we note that other QDs in the same sample show higher Purcell factors \cite{Tomm2021} than the QD used in these experiments -- so that the Purcell factor can be increased from 6.1 to 12 without any modifications to the cavity. By reducing optical losses we estimate that $\eta_{\textrm{optics}} \cdot \eta_{\textrm{coupler}}$ can be increased from 55.2\% to 90\%. Finally, single-photon detectors with quantum efficiency $\eta_{\textrm{detector}}=95$\% (instead of 82\%) are commercially available, and could also be used. These improvements would lead to $\eta=76$\% and would allow single-shot readout in less than 1\,ns with a readout fidelity of 99.5\%.

Although our readout speed is extremely fast, a key question is whether we can read the spin state fast enough to overcome the back-action in the Voigt geometry (in-plane magnetic field) as this is the configuration required for spin control. With $F_P = 12$, the branching ratio is 92.3\%. In the optimised case ($\eta= 76$\%), we expect we can achieve single-shot readout with a fidelity of 89.9\% below 1\,ns, while for our present experimental conditions, single-shot readout should already be possible with a fidelity as high as 77.4\% in 3\,ns. These readout fidelities are extremely promising. Our approach can thus overcome a key outstanding challenge, namely combining spin control and spin readout in one a single QD spin.

\bibliography{references}

\end{document}